\documentclass[twocolumn,twoside,slac_two]{revtex4}
\usepackage{graphicx}
\usepackage{fancyhdr}
\usepackage{hyperref}
\hypersetup{
    colorlinks=true,
    urlcolor=magenta,
}
\begin{document}

\title{Precision Measurement of Boron-to-Carbon ratio in Cosmic Rays from 2 GV to 2 TV with the Alpha Magnetic Spectrometer on the International Space Station.}

\author{V. Formato, on behalf of the AMS-02 collaboration}
\affiliation{INFN-Perugia, I-06100 Perugia, Italy}
\affiliation{CERN, Geneva, Switzerland}

\begin{abstract}
AMS-02 is a wide acceptance high-energy physics experiment installed on the International Space Station in May 2011 and it has been operating continuously since then. AMS-02 is able to separate cosmic rays light nuclei ($1\leq Z \leq 8$) with contaminations less than $10^{-3}$.
The ratio between the cosmic rays Boron and Carbon fluxes is known to be very sensitive to the properties of the propagation of cosmic rays in the Galaxy, being Boron a secondary product of spallation on the interstellar medium of heavier primary elements such as Carbon and Oxygen. A precise measurement reaching the TeV region can significantly help understanding cosmic rays propagation in the Galaxy and the amount of matter traversed before reaching Earth.
The status of the measurement of the boron-to-carbon ratio based on 10 millions Boron and Carbon events is presented.
\end{abstract}

\maketitle

\thispagestyle{fancy}

\section{Introduction}
Primary cosmic ray (CR) species are those present at the site of CR acceleration and that get then injected into the Galaxy where they are scattered randomly by irregularities in the galactic magnetic field. Secondary species are elements that are not present at the acceleration sites but are originated as by-products of reactions of heavier primary elements with the interstellar medium. Due to the fact that secondary CRs are created exclusively during the propagation of primary CRs, secondary-to-primary flux ratios offer a direct probe of the mechanisms that rule propagation of cosmic rays in the Galaxy \cite{2007ARNPS..57..285S,2015PhRvD..92h1301T,2015ApJ...803L..15T}.

Among all the possible secondary-to-primary cosmic rays flux ratios the Boron to Carbon ratio (B/C) is the most convenient secondary-to-primary ratio to be measured experimentally since Boron and Carbon are neighbouring elements in the table of elements and have similar interactions in matter and it doesn't require mass separation capabilities by the experimental apparatus. Measurements from the seventies up to today are available in the following Refs.
\cite{1978ApJ...226.1147O,1987ApJ...322..981D,1980ApJ...239..712S,1990A&A...233...96E,1985ICRC....2...16W,1991ApJ...374..356M,1994ApJ...429..736B,2010ApJ...724..329A,2008ICRC....2....3P,2008APh....30..133A,2012ApJ...752...69O,2014ApJ...791...93A}.

\begin{figure*}[ht]
\centering
\includegraphics[width=\textwidth]{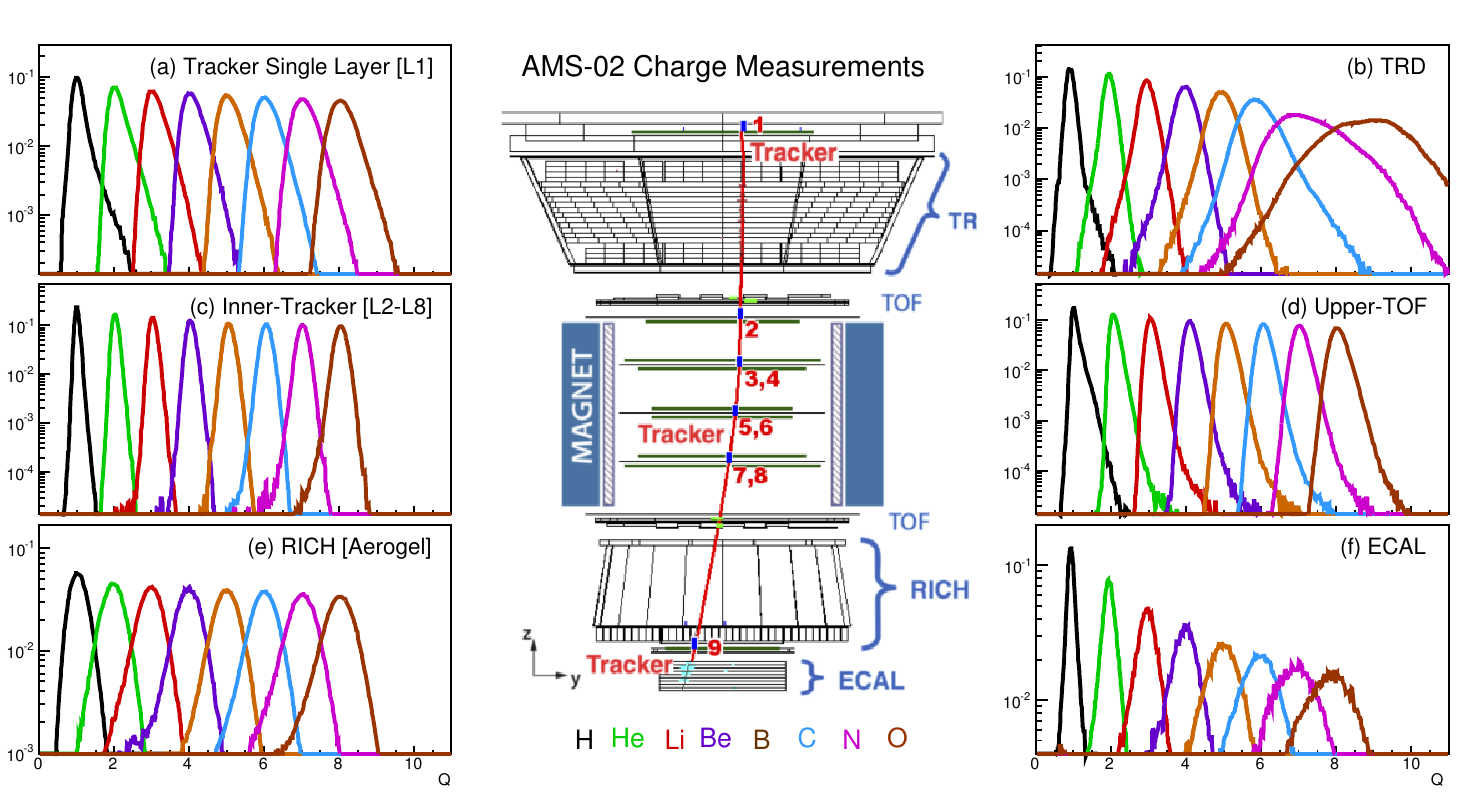}
\caption{A charged particle traversing AMS-02 is depicted in the center of the figure. The absolute charge $|Z|$ of the particle is measured several times along the particle trajectory. From the top to the bottom of the detector charge estimators are given by: Tracker L1, TRD, Upper TOF, Inner Tracker, Lower TOF, RICH, Tracker L9, and ECAL. On the left and on the right the charge distributions of the listed measurements are presented for clean samples of ions with $1 \leq |Z| \leq 8$.} \label{fig:charge_meas}
\end{figure*}

AMS-02 is a general purpose high energy magnetic spectrometer in space. The design of the AMS-02 detector is described in detail in Ref. \cite{2012IJMPE..2130005K}. It consists of nine planes of precision Silicon Tracker, a transition radiation detector (TRD), four planes of time-of-flight counters (TOF), a permanent magnet, an array of anti-coincidence counters (ACC) surrounding the inner tracker, a ring imaging Cherenkov detector (RICH), and an electromagnetic calorimeter (ECAL).
Several sub-detectors offer a measurement of the charge of the particle along its trajectory: on top of AMS by a layer of Silicon sensors constituting the Tracker L1; in the TRD by the combination of measurements in the 20 layers of straw tubes; in the Inner Tracker by the combination of 7 single layer Tracker measurements (from L2 to L8); in the Upper TOF (UTOF) by the combination of the measurements of two layers of scintillating counters; on the Lower TOF (LTOF) similarly to UTOF; in the RICH by counting the number of \c Cerenkov photons emitted by the particle inside the radiator; on the Tracker L9; and by the the energy deposit measurement in the first layers of ECAL that show no inelastic interaction. In Fig. \ref{fig:charge_meas} the charge distribution of the multiple AMS charge estimators are presented.
Best responses are given by the Inner Tracker, UTOF and LTOF with resolutions of $\Delta Z = 0.12$, $0.16$ and $0.16$ charge units respectively for $Z = 6$.

The measurement of the particle curvature in the magnetic field provides an estimate of the particle rigidity ($R = pc/Ze$, \emph{i.e.} momentum per unit charge). The nine position measurements along the particle trajectory on a lever arm of about 3 meters between L1 and L9 correspond to a Maximum Detectable Rigidity (MDR, that is, the rigidity for which $\Delta R / R = 1$) of about 2.6 TV for $Z = 6$ and 3.0 TV for $Z=5$.

Monte Carlo simulated events were produced using a dedicated program developed by the collaboration based on the {\tt GEANT-4.10.1} package \cite{2003NIMPA.506..250A}, where nucleus-nucleus interactions were modelled with the {\tt DPMJET-II.5} package \cite{1995PhRvD..51...64R} above 5 GeV/nucleon, and the Binary Light Ion Cascade model \cite{2003NIMPA.506..250A} for energy below the same threshold. The program simulates transport of ions through the materials of the AMS sub-detectors. The digitization of the signals is simulated according to the measured characteristics and response of the electronics.
The use of Monte Carlo in this analysis is limited to the estimation of geometric acceptances, evaluation of the response function of the Tracker, and the top-of-the-instrument correction.

\section{Selection}
Data collected in the first 60 months of AMS operation, corresponding to a collection time of $1.23 \times 10^8$ s, have been analysed, using only those seconds during which the detector was in normal operating conditions, AMS was pointing within $40^\circ$ of the local zenith, and the ISS was outside of the South Atlantic Anomaly. Downward going particles with charge 5 and 6 in the Inner Tracker and UTOF are firstly selected. The efficiency of this selection is very high, up to 98\% with a negligible probability of charge mis-identification. To exclude as many events with inelastic interaction in the detector as possible, only events that have a consistent charge measurement on Tracker L1 are used in the analysis. In order to have the best resolution at the highest rigidities, further selections are made by requiring track fitting quality criteria such as minimum request on $\chi^2$/d.f. in the bending coordinate. To select only primary CRs the measured rigidity is required to be greater than a factor of 1.2 times the maximum geomagnetic cutoff within the AMS field of view. The cutoff was calculated by backtracing particles \cite{2000PhLB..484...10A} from the top of AMS out to 50 Earth’s radii using the most recent IGRF \cite{2010GeoJI.183.1216F} geomagnetic model. The final sample consists in about $8.3 \times 10^{6}$ Carbons and $2.3 \times 10^6$ Boron \cite{2016PhRvL.117w1102A}.

These events are then used for 2 different analyses. An analysis that requires the presence of a charge measurement compatible with $Z = 5$, $6$ on the Tracker L9, and an analysis that does not have any additional request on the lower part of AMS. The first analysis has the largest level arm and exploits the best rigidity resolution, the second analysis instead has a much larger acceptance and lower interaction levels inside the detector. The two analyses have been found to be compatible and combined in the final B/C ratio to have, whenever possible, maximum statistics and the best rigidity resolution.

\begin{figure}[ht]
\centering
\includegraphics[width=\columnwidth]{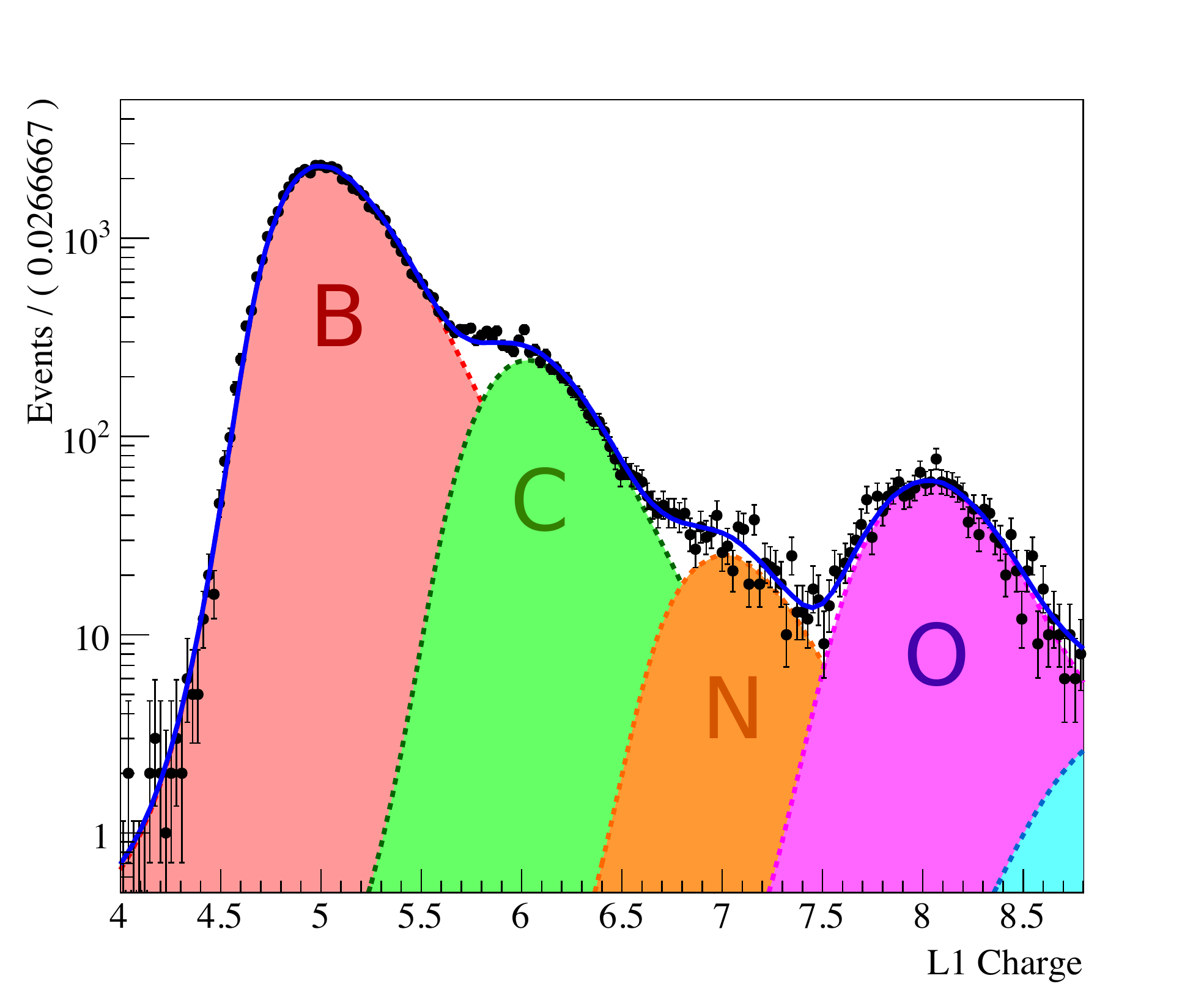}
\caption{Charge of the Tracker L1 for a low contamination Boron sample selected with in the Inner Tracker. The charge distribution shows a large population of not-interacting Boron events, as well as a population of higher $Z$ nuclei (C, N, O) that interacted in the upper part of AMS giving emerging B fragments. 
} \label{fig:L1charge}
\end{figure}

Nuclei may interact on AMS materials and split into fragments of lower charge. Fig. \ref{fig:L1charge} shows the charge distribution on top of AMS measured by L1 for a clean selection of Boron obtained with the UTOF and the Inner Tracker. The extra populations of Carbon, Nitrogen and Oxygen are due to the charge-changing processes O $\rightarrow$ B, N $\rightarrow$ B and C $\rightarrow$ B happening between L1 and UTOF. To estimate the purity of the sample on L1 (after the corresponding charge selection is applied) a fitting procedure has been developed. Reference spectra for each charge were derived and used to fit the distribution obtained after cuts, as presented in Fig. \ref{fig:L1charge}. 
After cutting on the L1 charge the estimated purity of the Boron sample is $> 97\%$ with an efficiency of $\sim 96\%$, while the Carbon sample purity and efficiency are very close to 1, since the large abundance of Carbon allows selections with high efficiency and negligible backgrounds.

Charge-changing process happening in materials above L1 produce an additional source of irreducible background, which has been estimated from simulation, using MC samples generated according to AMS flux measurements. This determination relies in the MC simulation of materials above L1, mainly  supporting structures, and is validated by comparing the Inner Tracker charge distribution for $Z=6$ events as determined by the L1 in both data and MC \cite{2016PhRvL.117w1102A}. 
This irreducible background in the boron sample amounts to 2\% at 2 GV and increases up to 8\% at 2.6 TV, while for the carbon sample it is below 0.5\% over the entire rigidity range.

The bin-to-bin migration of events was corrected using the unfolding procedure described in Ref. \cite{2015PhRvL.114q1103A}.

\section{Results}
The Boron-to-Carbon ratio can be written as a ratio of two isotropic flux $\Phi^Z_i$ for the $i$\textsuperscript{th} rigidity bin $(R_i,R_i +\Delta R_i)$ as:
$$
\Phi^Z_i = \frac{N^Z_i}{A^Z_i \varepsilon^Z_i T_i \Delta R_i}
$$
where $N_i^Z$ are the number of events corrected for charge migrations inside the detector, for charge migration above L1 and for the rigidity resolution function. $A^Z_i$ is the geometric acceptance evaluated in MC, $\varepsilon_i^Z$ is the byproduct of efficiencies estimated directly from data, $T_i$ is the collection time. With this definition, the B/C ratio can be expressed as:
$$
\text{B}/\text{C} = \frac{ \Phi^{\text{B}}_i }{ \Phi^{\text{C}}_i } = \frac{N^{\text{B}}_i}{N^{\text{C}}_i}
\cdot
\left[ \frac{A^{\text{B}}_i \varepsilon^{\text{B}}_i}{A^{\text{C}}_i \varepsilon^{\text{C}}_i} \right]^{-1}
$$
In the B/C ratios of efficiency terms tends to cancel out since the interaction of Boron and Carbon in matter are similar. The ratio of the efficiencies is estimated directly from data and includes the trigger efficiency, TOF efficiency, tracking efficiency, and the efficiency of finding hits on the external layers of the Tracker. To validate the MC predictions, Boron and Carbon event samples that cross the materials between L8 and L9 (Lower TOF and RICH) and reach L9 without interacting are used. The fraction of surviving events measured in data is compared with MC calculations with Glauber-Gribov inelastic cross sections varied within $\pm 10\%$. The resulting cross section with the best agreement to data above 30 GV were chosen \cite{2016PhRvL.117w1102A}.

\begin{figure}[ht]
\centering
\includegraphics[width=\columnwidth]{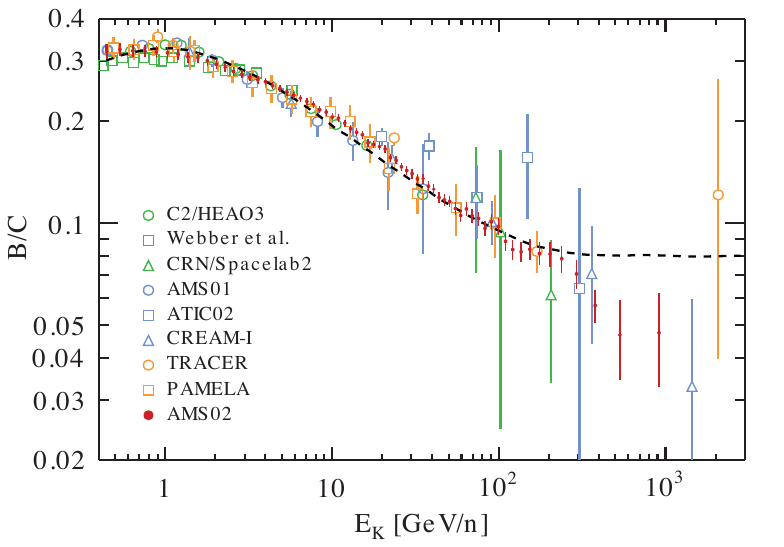}
\caption{The boron to carbon ratio as a function of kinetic energy per nucleon $E_k$ compared with previous measurements \cite{1978ApJ...226.1147O,1987ApJ...322..981D,1980ApJ...239..712S,1990A&A...233...96E,1985ICRC....2...16W,1991ApJ...374..356M,1994ApJ...429..736B,2010ApJ...724..329A,2008ICRC....2....3P,2008APh....30..133A,2012ApJ...752...69O,2014ApJ...791...93A}. The dashed line is the B/C ratio required for the model of Ref. \cite{2014ApJ...786..124C}.
} \label{fig:BCekin}
\end{figure}

The derived B/C ratio in the rigidity range between 1.9 GV to 2.6 TV is presented in Fig. \ref{fig:BCekin} compared with previous results \cite{1978ApJ...226.1147O,1987ApJ...322..981D,1980ApJ...239..712S,1990A&A...233...96E,1985ICRC....2...16W,1991ApJ...374..356M,1994ApJ...429..736B,2010ApJ...724..329A,2008ICRC....2....3P,2008APh....30..133A,2012ApJ...752...69O,2014ApJ...791...93A}. Errors include both statistics and systematics. The main source of error above 50 GV is due to the statistics of both the Boron and Carbon samples, while systematics uncertainties account for charge migration inside AMS, charge migration above L1, rigidity migration, efficiency and acceptance ratio corrections.

To compare with previous result, published mostly in kinetic energy per nucleon, the rigidity measurement was converted into kinetic energy, the details of such conversion can be found in Ref. \cite{2016PhRvL.117w1102A}.
\section{Conclusions}
The light nuclei cosmic ray Boron to Carbon flux ratio is very well known sensitive observable for the understanding of the propagation of cosmic rays in the Galaxy, being Boron a secondary product of spallation on the interstellar medium of heavier primary elements such as Carbon and Oxygen. A precision measurement based on 10 million events of the Boron to Carbon ratio in the rigidity range from 1.9 GV to 2.6 TV has been presented.


\end{document}